\documentclass[twocolumn]{aastex6}
\usepackage{amsmath,amssymb,bm, graphicx,xspace,multirow}
\usepackage{epsfig}
\usepackage{color}
\RequirePackage{lineno}
\usepackage{txfonts}
\usepackage{float}
\usepackage[font=footnotesize]{caption}

\usepackage{color}

\def\fermi{{\it Fermi}-LAT\xspace}

\def\aap{Astronomy and Astrophysics}

\def\apjl{ApJL}

\def\apjs{ApJS}

\begin{document}

\title{Origin of Galactic sub-PeV diffuse gamma-ray emission: constraints from high-energy neutrino observations}
\author{Ruo-Yu Liu$^{1,2}$ and Xiang-Yu Wang$^{1,2}$}

\affil{$^1$School of Astronomy and Space Science, Xianlin Road 163, Nanjing University, Nanjing
210023, China; Emails: {\color{blue} ryliu@nju.edu.cn; xywang@nju.edu.cn}\\
$^2$Key laboratory of Modern Astronomy and Astrophysics (Nanjing University), Ministry of Education, Nanjing 210023, People's Republic of China\\}

\begin{abstract}
Very recently, diffuse gamma rays with $0.1\,{\rm PeV}<E_\gamma <1\,$PeV have been discovered from the Galactic disk by the Tibet air shower array and muon detector array (Tibet AS+MD array). While the measured sub-PeV flux may be compatible with the hadronic origin in the conventional Galactic cosmic ray propagation model, we find that it is in possible tension with the non-detection of Galactic neutrino emissions by the IceCube neutrino telescope. We further find that the presence of an extra cosmic ray component of relatively hard spectrum, which is probably related to the Cygnus Cocoon region and other PeV cosmic-ray sources in the Galactic disk, would alleviate the tension. This scenario {implies} the existence of an extreme accelerator of {either protons or electrons} beyond PeV in the Cygnus region, {and predicts the continuation of the gamma-ray spectrum of Cygnus Cocoon up to 1\,PeV with a possible hardening beyond $\sim 30-100\,$TeV.} 
\end{abstract}

\maketitle

\section{Introduction}
The gamma-ray sky is dominated by the diffuse emission from the Galactic plane. The survey by gamma-ray satellites such as \textit{SAS}-2 \citep{Fichtel75}, EGRET on the \textit{Compton Gamma-Ray Observatory} \citep{Hunter97}, \textit{Fermi} Large Area Telescopes \citep[\fermi,][]{Fermi12_DGE, Neronov20} have measured diffuse gamma-ray emission from the Galactic plane from several tens of MeV up to TeV energies. Ground-based gamma-ray instruments can also measure Galactic diffuse gamma rays from a fraction of the Galactic plane due to the limited observable sky. Milagro \citep{Milagro05} and ARGO-YBJ \citep{ARGO15_diffuse} have extended the Galactic diffuse gamma-ray spectrum up to several TeV. The leading radiation mechanism for the diffuse Galactic gamma-ray emission (DGE) is believed to be the decay of neutral pions generated from hadronic interactions of cosmic ray (CR) protons with the interstellar medium \citep[e.g.][]{Dermer86, Masaki97, Strong10}, whereas energetic CR electron/positron pairs escaping from pulsar wind nebulae around middle-aged pulsars may also have an important contribution to the DGE at TeV band \citep{Linden18}. Therefore, the DGE can serve as a probe of the CR distribution in the Galactic disk, and be used to study the CR propagation and their origin. Very recently, \citet{Asgamma21_diffuse} reported detection of $0.1-1$\,PeV DGE by the Tibet AS+MD array. {All gamma rays above 398\,TeV are observed apart from $0.5^\circ$ of any known TeV sources and hence unlikely originate from leptonic sources since high-energy electrons cannot propagate far from sources before being cooled.} The detection of diffuse sub-PeV gamma-ray emission of the hadronic origin provides a good opportunity to study the origin of PeV CRs.

Since high-energy neutrinos always accompany with the production of pionic gamma rays, a diffuse high-energy neutrino background is expected from the Galactic plane. Recently, the IceCube neutrino telescope has obtained an upper limit on the neutrino flux of  the Galactic plane in the energy range of $1-500\,$TeV  \citet{IC17_GP}. This motivates us to compare the diffuse gamma-ray emission measured by the Tibet AS+MD array and the neutrino flux constraint from IceCube.  As will be shown in this Letter, we find that if we require the model neutrino intensity to be consistent with the 90\% C.L. upper limit of Galaxy, the model gamma-ray intensity is lower than the measured flux by the Tibet AS+MD array especially at the highest-energy bin in $398-1000\,$TeV. It may indicate that some additional sources contribute to the diffuse sub-PeV gamma-rays measured by the Tibet AS+MD array. {For example, \citet{Ahlers14} suggested that remnants of historical yet unresolved hypernovae, which are believed to be capable of energizing protons up to 1\,EeV \citep{Wang07,Budnik08}, may contribute to the sub-PeV DGE while not overproduce Galactic neutrinos.}

Interestingly, as pointed out by \citet{Asgamma21_diffuse},  4 out of 10 measured events above 398\,TeV are detected within $4^\circ$ from the center of the so-called Cygnus Cocoon. Therefore, a non-negligible fraction of the diffuse gamma-ray flux measured by the Tibet AS+MD array may originate from the Cygnus region. Indeed, various instruments have detected excess of gamma-ray emission from GeV up to 100\,TeV from Cygnus Cocoon \citep{Fermi11_cygnus, ARGO12_cygnus, VERITAS18_cygnus, HAWC21_cygnus}, which is probably a superbubble related to the massive star cluster Cygnus OB2 and is also in spatial coincidence with a pulsar wind nebula, PWN TeV~J2032+4130 and a supernova remnant, $\gamma$ Cygni. All of these objects are possible sources of energetic protons and electrons so that they are potential sources of these gamma rays. We speculate that a significant fraction of sub-PeV gamma-rays detected by the Tibet AS+MD array may originate from the Cygnus region and/or some other extended sources in the Galactic disk, rather than the true diffuse gamma rays.  We will study whether this scenario can alleviate the tension.

\section{Gamma-ray and Neutrino Emission of Cosmic Rays in the Galactic disk}
To calculate the diffuse gamma-ray and neutrino emission produced by the proton-proton ($pp$) collisions between CRs and the interstellar medium, we need to know their spatial distributions, i.e., $n_{\rm CR}(E_p,r,z)$ and $n_{\rm ISM}(r,z)$ with $E_p$ being the CR energy, $R$ the radius from the Galactic center projected in the Galactic plane and $z$ the height from the Galactic plane. We here employ the model developed by \citet{Lipari18}, which can successfully reproduce both the flux (with an error of order $10\%-20\%$ depending on the direction) and the main features of the angular distribution of the diffuse gamma-ray flux measured by \fermi. The model parameterizes $n_{\rm CR}(E_p,R,z)$ and $n_{\rm ISM}(R,z)$ as well as the CR slope as a function of $R$ and $z$ analytically so it can save a lot of computation time for modelling the CR transport and spatial distribution. Following the ``factorized model'' in that paper, we can then obtain the diffuse gamma-ray or neutrino intensity (in unit of $\rm GeV~cm^{-2}s^{-1}sr^{-1}$) in certain direction with Galactic coordinate $(l,b)$ by performing the line-of-sight integration
\begin{equation}
\begin{split}
I_{\gamma/\nu}(E_{\gamma/\nu}, l,b)&= \int \int \mathcal{F}_{\gamma/\nu}\left\{E_{\gamma/\nu}, n_{\rm CR}(E_p,r, l, b), n_{\rm ISM}(r,l,b) \right\}\\
&\times \exp\left[-\tau(E_\gamma,r)\right]dE_p dr
\end{split}
\end{equation}
where $\mathcal{F}_{\gamma/\nu}$ denotes the operator calculating the gamma-ray or neutrino emissivity of $pp$ collisions following the semi-analytical method developed by \citet{Kelner06}, $r$ is the distance of a certain place from Earth in the direction $(l,b)$ and the corresponding $R$ and $z$ can be found by $R=\left(R_E^2+(r\cos b)^2-2R_Er\cos b\cos l\right)^{1/2}$ and $z=r\sin b$, where $R_E=8.5\,$kpc is the distance of Earth from the Galactic center. $\tau(E_\gamma,r)$ is the opacity of the pair production of high-energy gamma-ray photon on the cosmic microwave background and the interstellar radiation field. The latter is based on the model by \citet{Popescu17}. Note that this term will not appear in the equation if we calculate the neutrino intensity. To compare with measured DGE spectra from certain region of the Galactic plane, we need to average the intensity over the region of interest, i.e.,
\begin{equation}
\bar{I}_\gamma=\Omega_\gamma^{-1}\int\int I_{\gamma}\sin (90^\circ-b)db dl.
\end{equation}
For example,  the DGE spectra shown in \citet{Asgamma21_diffuse} are extracted from $25^\circ <l <100^\circ$ $|b|<5^\circ$ (corresponding to $\Omega_\gamma=0.228\,$sr), and $50^\circ <l <200^\circ$ $|b|<5^\circ$ (corresponding to $\Omega_\gamma=0.456\,$sr) respectively. The neutrino analysis by IceCube uses the data including the entire Galactic plane, but is primarily sensitive to the northern hemisphere. Therefore, to compare with the neutrino upper limit, we only average the intensity over the Galactic plane in the northern hemisphere (or with declination $\delta \geq 0$), i.e.,
\begin{equation}
\bar{I}_\nu=\Omega_\nu^{-1}\int\int I_{\nu}\theta \left(\delta(l,b)\right)\sin (\pi/2-b)db dl,
\end{equation}
where $\theta$ is the Heaviside function and $\delta$ can be found by $\sin\delta=\sin(27.13^\circ)\sin b+\cos(27.13^\circ)\cos b\cos(122.93^\circ-l)$, where $27.13^\circ$ is the declination of the north Galactic pole and $122.93^\circ$ is the longitude of the northern equatorial pole. The corresponding solid angle of the region of interest for the neutrino emission is $\Omega_\nu=0.553\,$sr.

\begin{figure}
\centering
\includegraphics[width=1\columnwidth]{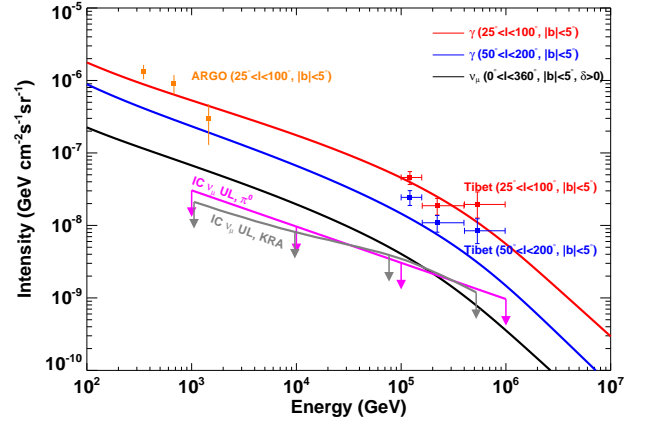}
\caption{Comparison between the diffuse gamma-ray and (anti-)muon neutrino intensities from the Galactic disk ($|b|<5^\circ$) expected by the model and observations. The solid red and blue curves show the average gamma-ray intensity in the region of $25^\circ<l<100^\circ$ and $50^\circ<l<200^\circ$ respectively, while the black curve shows the average neutrino intensity from the Galactic disk in the northern hemisphere $\delta>0$. The red and blue squares are the measured spectra of diffuse gamma rays from $25^\circ<l<100^\circ$ and $50^\circ<l<200^\circ$ respectively \citep{Asgamma21_diffuse}. The orange squares presents the measured spectra of diffuse gamma rays from $25^\circ<l<100^\circ$ by ARGO-YBJ \citep{ARGO15_diffuse}. The magenta bar with arrows exhibits the 90\% C.L. upper limit for $\nu_\mu+\bar{\nu}_\mu$ neutrino flux of the Galactic plane using the \fermi $\pi^0$-decay spatial template with assuming $E_\nu^{-2.5}$ \citep{IC17_GP}, while the grey bar with arrows is the same except for using the template predicted by the KRA model \citep{Gaggero15}.}
\label{fig:diffuse}
\end{figure}

We compare the gamma-ray intensity and neutrino intensity predicted by the model with the observations in Fig.~\ref{fig:diffuse}. For the predicted gamma-ray intensity, our result is the generally the same with that shown in Fig.20 of \citet{Lipari18}. The model intensity is slightly lower than the measured intensity by the Tibet AS+MD array, in particular at the highest-energy bin, for both $25^\circ<l<100^\circ$ and $50^\circ<l<200^\circ$. Such a difference is not significant considering the systematic error, as pointed out by \citet{Asgamma21_diffuse}. However, we also see that the accompanying neutrino intensity exceeds the 90\% confidence level (C.L.) upper limit of IceCube\footnote{The neutrino upper limit given in \citet{IC17_GP} is scaled to represent an all-sky integrated flux. We need to divide the given upper limit by a factor of $4\pi$ for comparison.} below 100\,TeV by about a factor of 2. In other words, there might be a tension between the diffuse sub-PeV gamma-ray data and the neutrino data, and if we want to reconcile the model neutrino intensity with the IceCube's upper limit, we need to multiply a factor of 0.5 to the neutrino intensity. In the mean time, it would reduce the predicted gamma-ray intensity by the same factor and make it insufficient to explain the DGE data especially for the highest-energy bin. Note that there are some uncertainties of the model, such as the metallicty of both CRs and the interstellar medium which would influence the predicted gamma-ray or neutrino intensity up to a factor of 2, as well as the spatial variation of CR density and spectra in the Galactic plane and etc. Therefore, it is acceptable to have the flexibility of a factor of a few for the model intensity.

{In fact, the model uncertainty influencing the amplitude of the gamma-ray intensity would not affect our result significantly because we re-scale the gamma-ray intensity according to the co-produced neutrino intensity and the IceCube upper limit. On the other hand, the spectral variation of CRs across the Galaxy is important. The factorized model of \citet{Lipari18} employed here considers the hardening of the CR spectrum towards the inner Galaxy, from a slope of 2.8 in the periphery of the Galaxy to 2.4 at the Galactic center. \citet{Lipari18} also constructed a ``no-factorized model'' with a uniform CR spectral slope, identical to the locally measured one, throughout the Galaxy. If we adopt the no-factorized model, the resulting diffuse gamma-ray spectrum would be softer than the present one for $25^\circ<l<100^\circ$. Consequently, it will be more insufficient to explain the measured intensity of the highest-energy bin solely with the diffusive CRs and strengthen our motivation to investigate additional contributions from extended sources. In contrast, if the CR spectral slope turns out to be harder than that in the factorized model in the inner Galaxy, the tension between the sub-PeV DGE and the Galactic neutrino flux upper limit can be alleviated.}

\section{Possible contribution from Cygnus Cocoon and other PeV CR sources}
It is reported that the total event number in the 398\,TeV -- 1000\,TeV bin observed by the Tibet AS+MD array is ten in each of the region $25^\circ < l < 100^\circ$ and $50^\circ < l < 200^\circ$. Among both of them, 4 out of 10 photons originate from the region within $4^\circ$ of the center of Cygnus Cocoon at $l\approx 80^\circ$ \citep{Asgamma21_diffuse}. {The Cygnus Cocoon region contains some potential sources for PeV CRs such as the massive star cluster Cygnus OB2 and the supernova remnant Gamma Cygni, so the energetic, fresh CRs may permeate throughout the entire region with a higher density than the average one in the interstellar medium.} Although the analysis of the diffuse emission has excluded the region within $0.5^\circ$ of the known TeV sources \citep{Asgamma21_diffuse}, the radial profile of the TeV gamma-ray of Cygnus Cocoon is quite extended which can be described by a Gaussian profile with width of $2.1^\circ$ \citep{HAWC21_cygnus}. Therefore, most of the emission of Cygnus Cocoon probably has been counted in the diffuse emission.
As a result, we should also consider an additional component from Cygnus Cocoon when modelling the intensity in both regions of $25^\circ < l < 100^\circ$ and $50^\circ < l < 200^\circ$.


The TeV spectrum of the Cygnus Cocoon region has been measured by HAWC up to 100\,TeV, which can be described by a power-law function with an index of 2.6 \citep{HAWC21_cygnus}. If part of the DGE between 398\,TeV -- 1000\,TeV measured by the Tibet AS+MD array can be attributed to Cygnus Cocoon, it implies that its spectrum should continue somehow up to 1\,PeV and indication the existence of an extreme accelerator of CRs in that region processing protons up to at least 10\,PeV. To evaluate the sub-PeV flux of Cygnus Cocoon, however, we need to know the relative exposure time of the instrument on the Cygnus region and on other region of the Galactic plane, which is not given. If we simply assume a uniform exposure of the Tibet AS+MD array over the observed Galactic plane, the fact that 4 out of 10 total events in both $25^\circ < l < 100^\circ$ and $50^\circ < l < 200^\circ$ from Cygnus Cocoon would mean that a fraction $f_{\rm cyg}\simeq 40\%$ of the total DGE flux ($\Omega_\gamma \bar{I}_\gamma$) at this energy originates from the source. Comparing it with the HAWC data, we see a flattening of the spectrum above $30-100\,$TeV (see the {filled} red and blue squares in Fig.~\ref{fig:cygnus}) and this would imply a different origin of the emission below 30\,TeV. We note that this is most likely an overestimation of the sub-PeV flux since the exposure for Cygnus Cocoon would be longer than the average value for the Galactic plane\footnote{Roughly speaking, the exposure time is larger if the source's declination is closer to the instrument's declination for a fixed live time. The declination of Cygnus Cocoon is $40^\circ$ while that of the Tibet AS+MD array is $30^\circ$, which is quite close to each other. So we expect the exposure time of Cygnus Cocoon is longer than the average of the Galactic plane.}, but we take it as a reference case and refer to it as CASE I. On the other hand, if we assume the flux of Cygnus Cocoon at 398\,TeV -- 1000\,TeV is consistent with the extrapolation of HAWC's spectrum, Cygnus Cocoon would account for only about $f_{\rm cyg}\simeq 5\%$ of the DGE intensity at the highest-energy bin measured by the Tibet AS+MD array (see the open red and blue squares in Fig.~\ref{fig:cygnus}). We refer to this case as CASE II.

\subsection{The Case of Hadronic Origin}
The HAWC observation on Cygnus Cocoon favors a hadronic origin of the TeV gamma-ray emission \citep{HAWC21_cygnus}. For a phenomenological modelling of the gamma-ray emission of Cygnus Cocoon, we consider a total proton energy of $W_p$ contained in the Cygnus region with a spectrum following $N_{p,\rm cyg}\equiv dN_p/dE_p=N_0E^{-2}\exp(-E_p/E_c)$ to explain the emission above 30\,TeV of Cygnus Cocoon in CASE I, and we assume the proton spectrum to be $N_{p,\rm cyg}\equiv dN_p/dE_p=N_0E_p^{-2}\left(1+E_p/{E_b}\right)^{-1}$ to account for also the spectrum below 1\,TeV in CASE II, where $N_0$ can be determined by $\int E_p(dN_p/dE_p)dE_p=W_p$. 

\begin{figure}
\centering
\includegraphics[width=1\columnwidth]{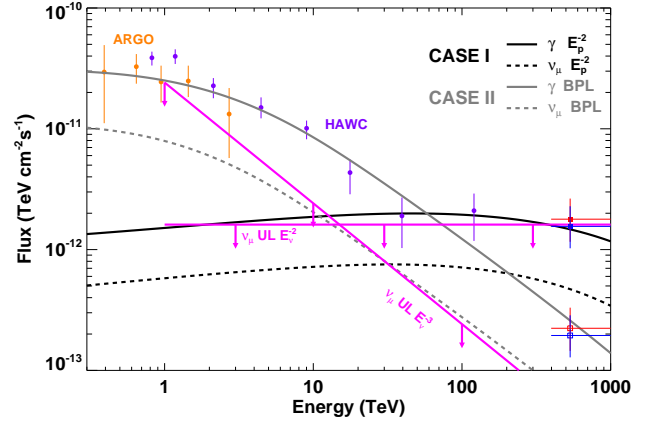}
\caption{TeV fluxes of gamma rays and (anti-)muon neutrinos from Cygnus Cocoon. The solid and dashed black curves show, respectively, the gamma-ray and neutrino flux in CASE I. The solid and dashed grey curves show the gamma-ray and neutrino flux in CASE II. Purple circles are the spectrum measured by HAWC \citep{HAWC21_cygnus} and orange circles are spectrum measured by ARGO-YBJ \citep{ARGO12_cygnus}. The filled red and blue squares show, respectively, the postulated flux of Cygnus Cocoon assuming that 40\% of the total flux in the 398\,TeV -- 1000\,TeV bin measured by the Tibet AS+MD array \citep{Asgamma21_diffuse} in $25^\circ<l<100^\circ$ and $50^\circ<l<200^\circ$ originate from the Cygnus region (CASE I), while the open red and blue squares assumes that 5\% of the total diffuse gamma-ray flux related to the Cygnus region (CASE II). The two magenta bars with arrows are the sum of the 90\% C.L. $\nu_\mu+\bar{\nu}_\mu$ upper limits for 2HWC~J2031+415 and Gamma Cygni assuming a $E_\nu^{-2}$-type spectrum \citep{IC20_ps} and an $E_\nu^{-3}$-type\protect\footnotemark[1] spectrum for comparison with the neutrino flux in CASE I and CASE II respectively.) }
\footnotetext[1]{\citet{IC20_ps} does not give the upper limit for $E_\nu^{-3}$ spectrum explicitly, but we observe from Fig.3 of the paper that the normalization of the flux at 1\,TeV for the $E_\nu^{-3}$-type spectrum is about 15 times higher than that for the $E_\nu^{-2}$-type spectrum.}
\label{fig:cygnus}
\end{figure}

We then can obtain the gamma-ray flux related to the sub-PeV DGE from Cygnus Cocoon by
\begin{equation}
\begin{split}
F_{\rm cyg, \gamma}(E_{\gamma})&=\frac{1}{4\pi r_{\rm cyg}^2}\int \mathcal{F}_{\gamma}\left\{E_{\gamma}, N_{p, \rm cyg}(E_p), n_{\rm cyg} \right\}\\
&\times \exp\left[-\tau(E_\gamma,r_{\rm cyg})\right]dE_p,
\end{split}
\end{equation}
given the average gas density of $n_{\rm cyg}=20\,\rm cm^{-3}$ in Cygnus Cocoon \citep{Aharonian19} and its distance $r_{\rm cyg}=1.4\,$kpc. We find the derived gamma-ray flux can match both the HAWC data and the postulated sub-PeV flux, with $E_b=20\,$TeV and $W_p=2.5\times 10^{49}\,$ergs in CASE I, and with $E_c=30\,$PeV and $W_p=1.5\times 10^{48}\,$ergs in CASE II as shown in Fig.~\ref{fig:cygnus}. {In this case, high-energy neutrinos are naturally expected from Cygnus Cocoon, as also predicted in some previous literature \citep[e.g.,][]{Beacom07, Evoli07, Bi09, Tchernin13, Fox13, Yoast-Hull17}.} We calculated the co-produced neutrino flux and find it consistent with the sum of the upper limits for 2HWC~J2031+415 and Gamma Cygni \citep{IC20_ps}, which are two potential CR sources related to Cygnus Cocoon. 

\begin{figure}
\centering
\includegraphics[width=1\columnwidth]{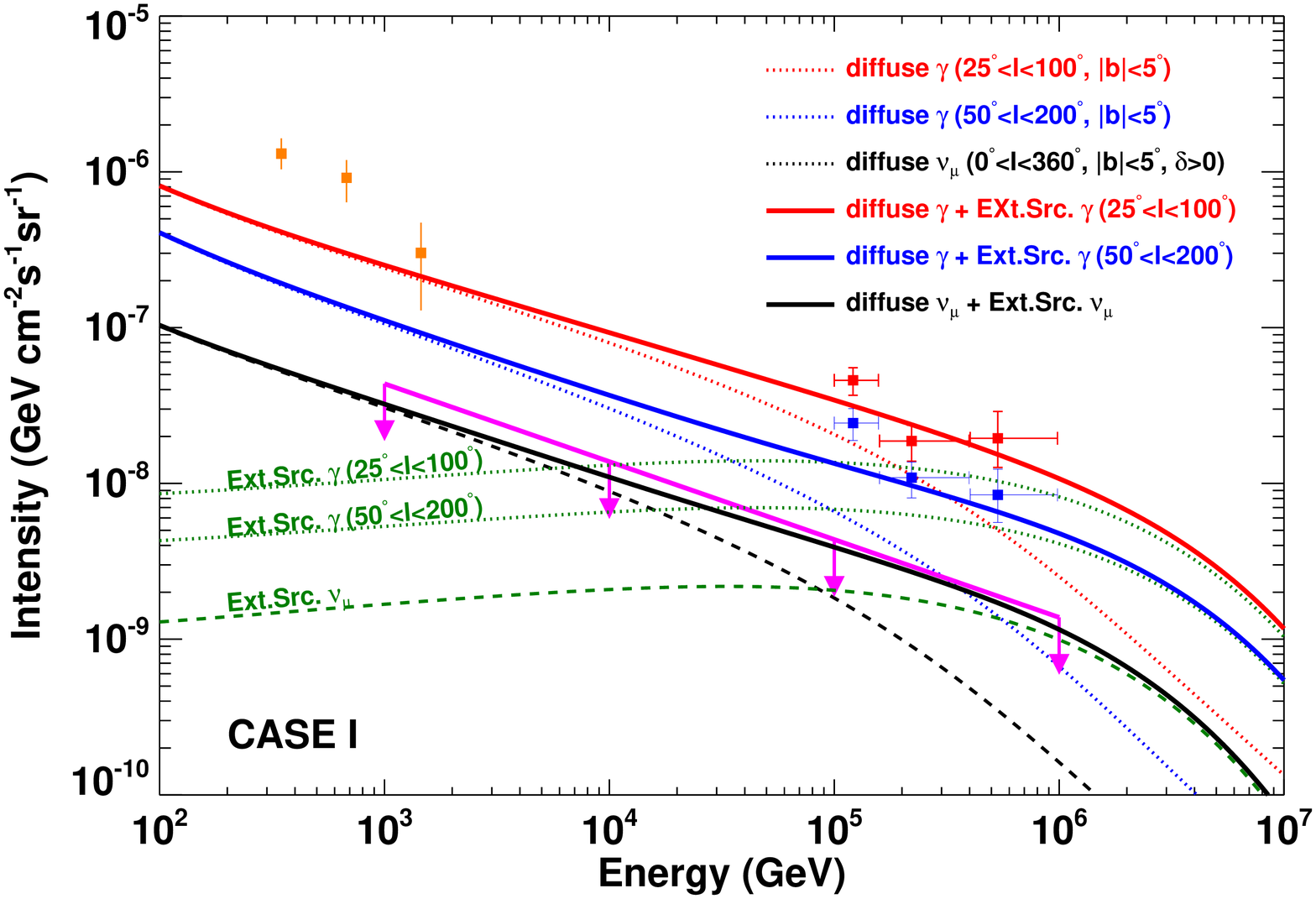}
\includegraphics[width=1\columnwidth]{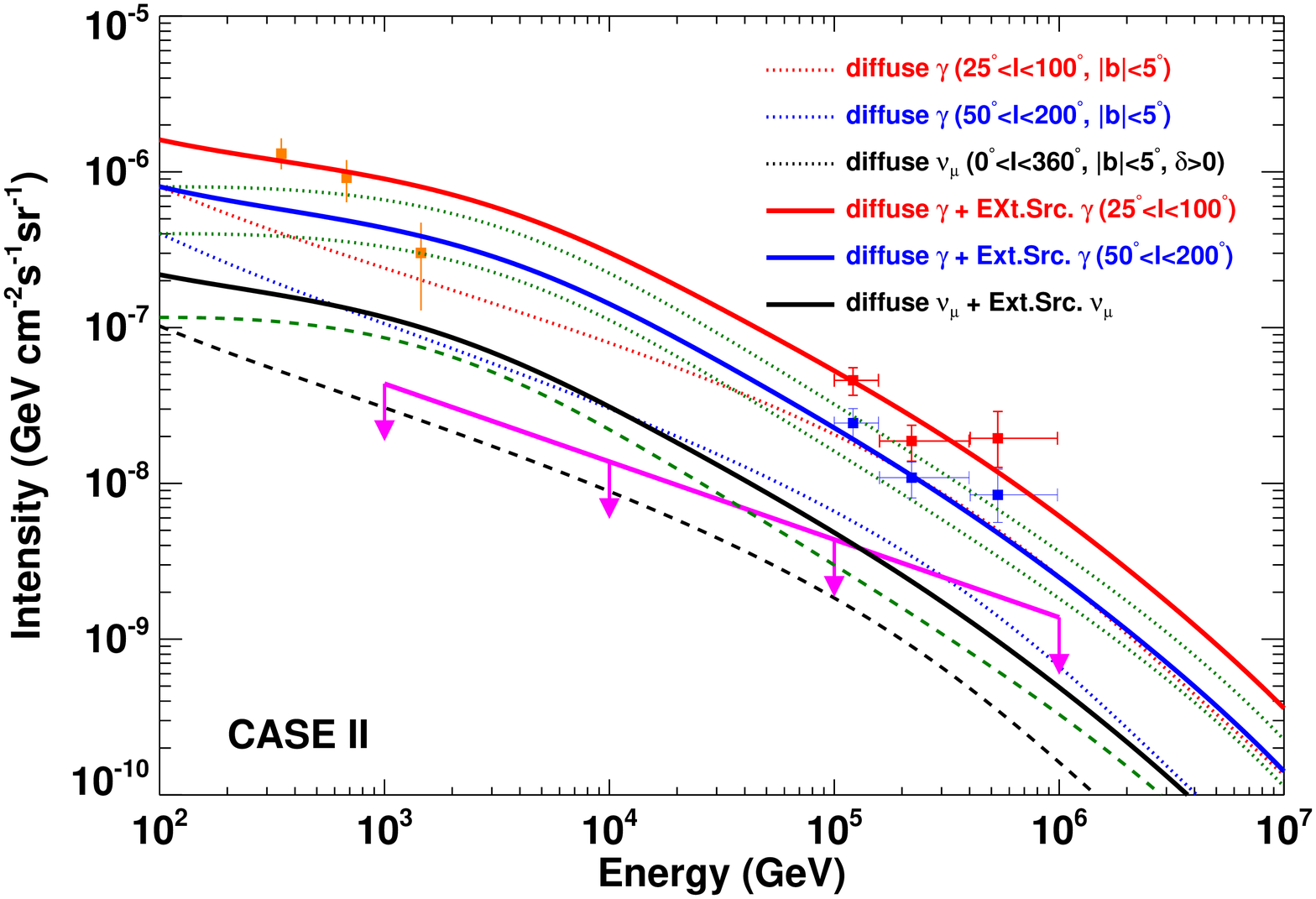}
\caption{Intensity of the diffuse gamma-ray and neutrino flux of the Galactic disk for CASE I (upper panel) and CASE II (lower panel). The  dotted red and blue curves are the model predicted gamma-ray intensities scaled by a factor of 0.45 for $25^\circ<l<100^\circ$ and $50^\circ<l<200^\circ$ respectively while the dashed black curve is the model predicted neutrino intensity of the northern hemisphere $\delta>0$ scaled by the same factor. The two dotted green curves show the gamma-ray contribution from extended CR sources represented by the emission of Cygnus Cocoon with being scaled up by a factor of $1.3$ and averaged over the solid angle corresponding to $25^\circ<l<100^\circ$ and $|b|<5^\circ$ (the upper one), and averaged over the solid angle corresponding to $50^\circ<l<200^\circ$ and $|b|<5^\circ$ (the lower one). The dashed green curve shows the corresponding neutrino intensity from the extended sources averaged over the solid angle corresponding to the Galactic disk in the northern hemisphere. The solid red, blue, and black curves are the sum of the gamma-ray intensity and neutrino from the Galactic disk and from the extended sources. The magenta bar with arrows show the 90\% C.L. $\nu_\mu+\bar{\nu}_\mu$ upper limit for both Galactic diffuse emission and source emission of five Galactic catalogs \citep{IC17_GP}. All the symbols have the same meaning with those in Fig.~\ref{fig:diffuse}.}
\label{fig:diffuse_cyg}
\end{figure}

In addition to Cygnus Cocoon, there may be other sources injecting CRs with energy beyond 1\,PeV into our Galaxy, although the others may be less powerful than Cygnus Cocoon. Similar to the case of Cygnus Cocoon, these super-PeV CRs diffuse quite fast, distributing over an extended region around their source, and produce gamma rays with a harder spectrum than that from the bulk of the interstellar medium. Without known the properties of these extended sources, we use Cygnus Cocoon's emission as a proxy of all these extended sources and parameterize the contribution of other sources by a coefficient $\xi$, i.e.,
\begin{equation}\label{eq:dge_source}
I_{\rm Ext.Src, \gamma}(E_{\gamma})=\Omega_{\gamma}^{-1}(1+\xi) F_{\rm cyg, \gamma}(E_\gamma).
\end{equation}
The result is shown in Fig.~\ref{fig:diffuse_cyg}. In CASE I, it needs $\xi=0.5$ to make the sum of the contribution by diffuse CRs and extended sources fit the diffuse gamma-ray spectra measured by the Tibet AS+MD array. The co-produced neutrino flux is consistent with the neutrino upper limit. Note that neutrinos now come from both discrete sources and diffusive CRs in the Galactic disk, so we compare it with the sum of the neutrino upper limits for diffuse Galactic emission as shown in Fig.~\ref{fig:diffuse} and the sources. For the latter, we employ the results in \citet{IC17_GP} for five catalogs of potential Galactic high-energy CR sources and sum up the upper limits of all the catalogs. In CASE II, since the contribution of Cygnus Cocoon is low, it needs $\xi=5$ to fit the diffuse gamma-ray spectra. However, due to the soft proton spectrum in this case, the co-produced neutrino flux exceeds the upper limit. Therefore, the value of $f_{\rm cyg}$, the fraction of the sub-PeV DGE that can be attributed to Cygnus region, is probably larger than 5\% (but lower than 40\% as CASE I is most likely an overestimation). With a larger $f_{\rm cyg}$, a harder proton spectrum will be inferred and subsequently the neutrino flux at low energies can be reduced. A larger $f_{\rm cyg}$ also implies that a hardening in the gamma-ray spectrum of Cygnus Cocoon would appear beyond $30-100$\,TeV, and this might suggest a different origin for the TeV emission below 30\,TeV. Note that, the spectra of other sources are not necessarily the same with that of Cygnus Cocoon. Thus, given a soft spectrum of Cygnus Cocoon with a low flux at sub-PeV such as in CASE II, the DGE at sub-PeV energy could be still explained if other sources have harder spectra.

\subsection{The Case of Leptonic Origin}\label{sec:lep}
{Although HAWC's observation on Cygnus Cocoon favors a hadronic origin, it does not exclude a leptonic origin completely for TeV gamma-ray emission above 1\,TeV. Indeed, if there exists a powerful CR proton accelerator in the center of Cygnus Cocoon, it should be able to produce high-energy electrons too. The pulsar wind nebulae powered by PSR~J2032+4127 and/or other unresolved pulsars may also inject high-energy electrons into ambient medium. Therefore, we also discuss a possible source contribution of leptonic origin to the sub-PeV DGE and we refer to this case as CASE III.

The sub-PeV gamma rays in the case of the leptonic origin can arise from the inverse Compton scattering (IC) of PeV electrons off the CMB radiation, whereas the IC process off radiation field of higher temperature such as the dust infrared radiation and the stellar optical-UV radiation are severely suppressed by the Klein-Nishina (KN) effect. In fact, even for CMB radiation with typical energy of $6\times 10^{-4}\,$eV, their scatterings with PeV electrons have also entered the KN regime, and thus we may expect the upscattered photon energy to be close to the electron energy, i.e., $E_\gamma \simeq E_e$. Electrons lose their energies as they radiate. The dominant cooling channel of electrons in the considered scenario is the synchrotron radiation with a cooling timescale of $t_{\rm syn}\simeq 500(E_e/{\rm 1\,PeV})^{-1}(B/5\,\mu {\rm G})^{-2}\,$yr, as long as the magnetic field in Cygnus Cocoon is not much weaker than the typical interstellar magnetic field $3-5\,\mu$G (compare the synchrotron cooling timescale $t_{\rm syn}$ and the IC cooling timescale $t_{\rm IC}$ in Fig.~\ref{fig:tcool}). Since the cooling timescale is much shorter than the dynamical timescale of the system, the IC spectrum can be estimated by $E_\gamma^2N_\gamma=E_e^2Q_{e,\rm inj}t_{\rm syn}/t_{\rm IC}\propto E_e^{1-p}\propto E_\gamma^{1-p}$. Here, $Q_{e,\rm inj}\propto E_e^{-p}$ is the injection spectrum, $t_{\rm syn}\propto E_e^{-1}$  and $t_{\rm IC}\propto E_e^0$ approximately in the range of $E_e=0.1-1\,$PeV. Thus, we obtain $N_\gamma \propto E_\gamma^{-p-1}$. To explain the DGE measured by the Tibet AS+MD array, the spectrum should be harder than $E_\gamma^{-2.7}$, because otherwise the source contribution would overshoot the measured DGE intensity at $0.1-0.4\,$PeV while it fits the $0.4-1\,$PeV intensity. This leads to a requirement of a hard electron injection spectrum with $p<1.7$.

On the other hand, a major difference of the leptonic case from the hadronic case is that electrons cool efficiently as they propagate while protons almost do not cool. Because the analysis of the Tibet AS+MD array masked the region within $0.5^\circ$ of each known TeV sources, those PeV electrons should at least diffuse to the region beyond $0.5^\circ$ of the source, which corresponds to a radius of 12\,pc at the nominal distance of 1.4\,kpc for Cygnus Cocoon. It then puts a constraint on the diffusion coefficient by $\sqrt{4Dt_{\rm syn}}\gg 12\,$pc, or $D(1\,{\rm PeV})\gg 2\times 10^{28}(B/5\mu\rm G)^2 cm^2s^{-1}$. This translates to $\beta\gg 0.015(B/5\,\mu{\rm G})^2$, if we follow \citet{HAWC21_cygnus} which considers an average diffusion coefficient in the Galaxy to be $D_{\rm ISM}(E)=3\times 10^{28} (E/10\,{\rm GeV})^{1/3}{\rm cm^2s^{-1}}$ and denote the diffusion coefficient in the Cygnus region by  $D_{\rm cyg}(E)=\beta D_{\rm ISM}(E)$. For a stronger magnetic field, a faster diffusion or a larger $\beta$ will be needed. We show an example of the expected DGE and SED of Cygnus Cocoon under CASE III in Fig.~\ref{fig:case3}, where we employ $p=1.5$, $\beta=0.1$, and an electron luminosity of $L_{e,\rm inj}=10^{35}\rm erg~s^{-1}$. Note that high-energy neutrinos are not expected from the sources in this case. We see that the DGE measured by the Tibet AS+MD array is marginally explained although the model overpredicts the 200\,TeV DGE intensity in $25^\circ <l <100^\circ$.

\begin{figure}
\centering
\includegraphics[width=1\columnwidth]{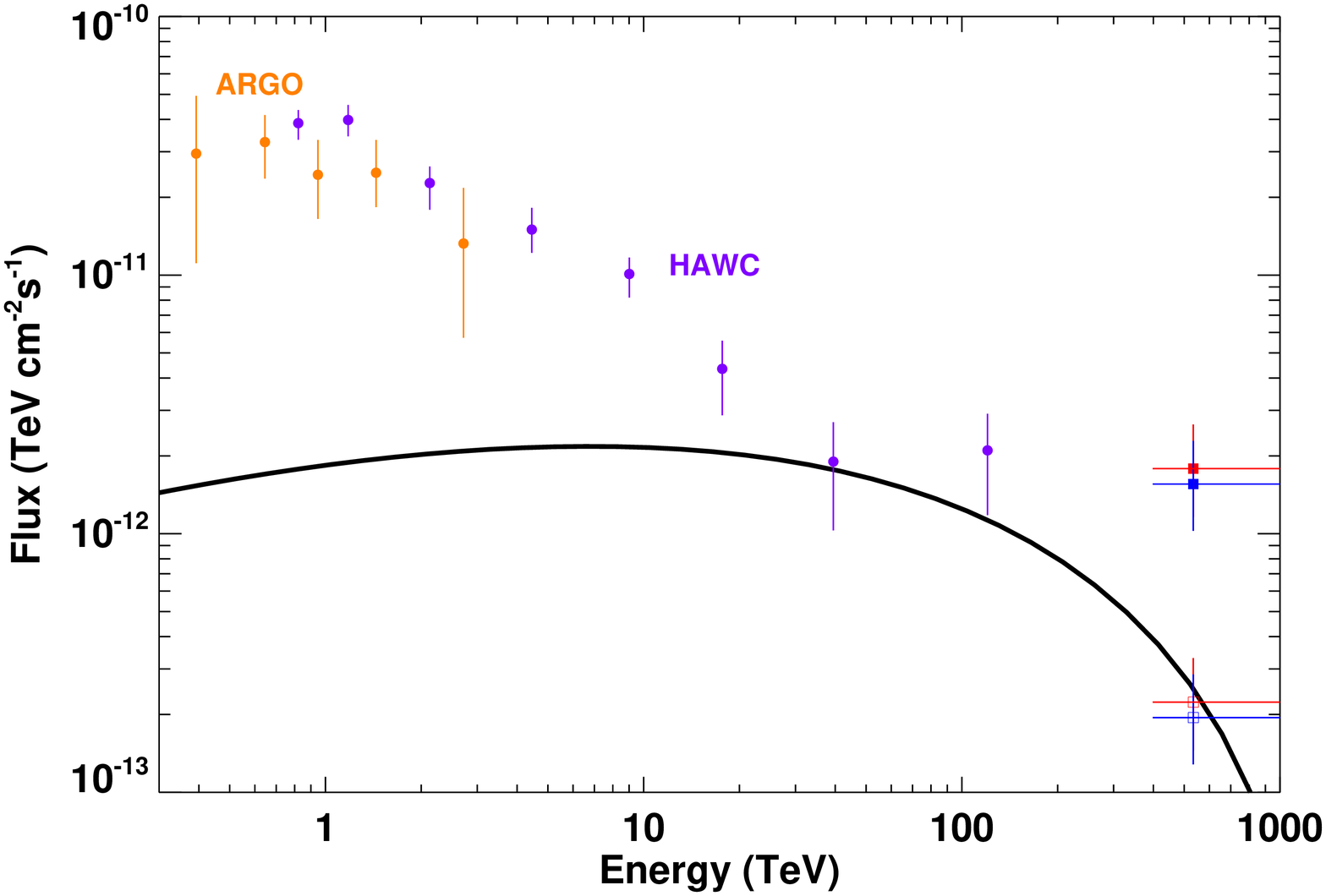}
\includegraphics[width=1\columnwidth]{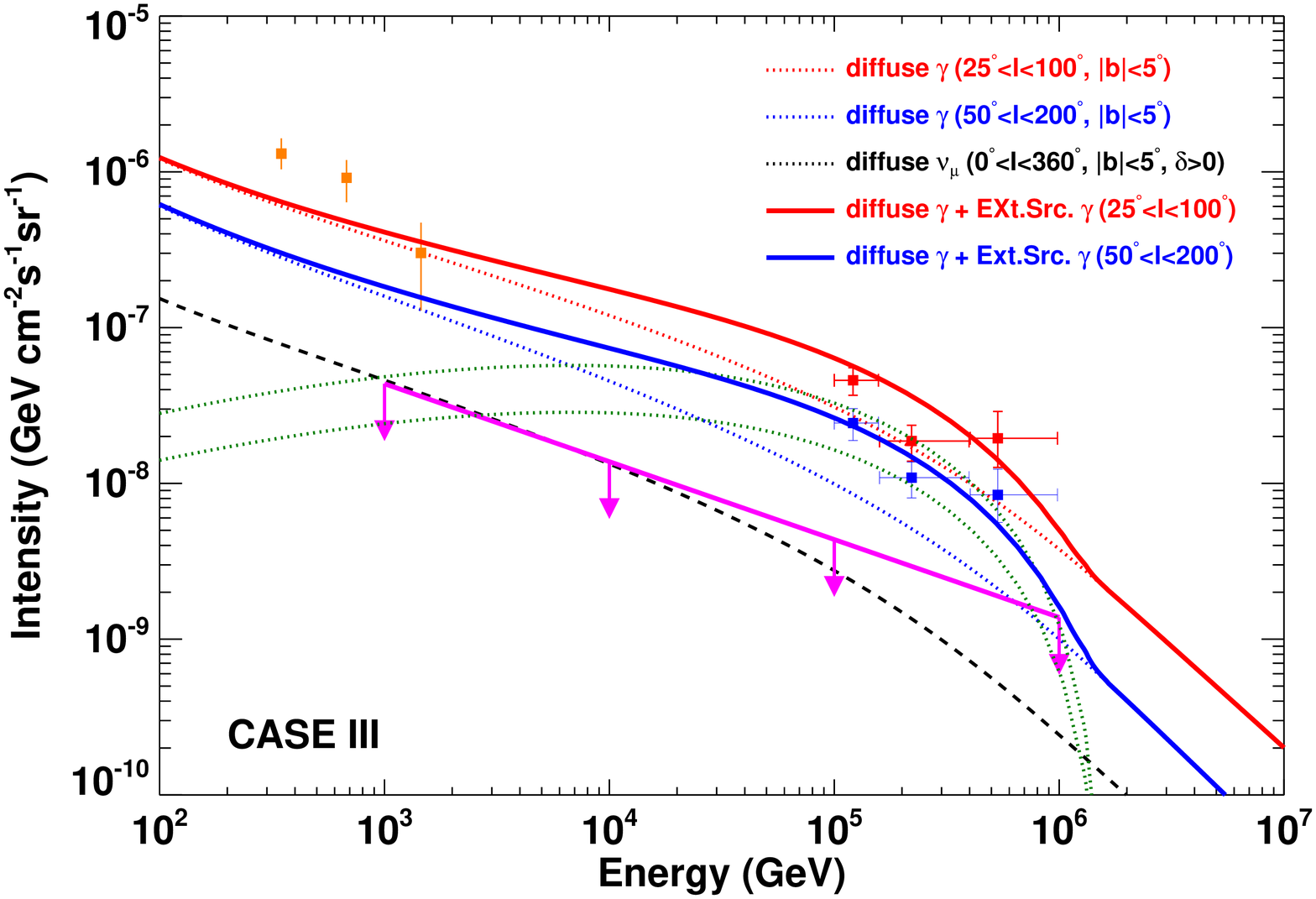}
\caption{{\bf Upper}: Same as Fig.~\ref{fig:cygnus} but for CASE III (the leptonic source contribution). 
{\bf Lower}: Same as Fig.~\ref{fig:diffuse_cyg} but for CASE III . $p=1.5$, $\beta=0.1$, $\xi=5$ and $L_{e,\rm inj}=10^{35}\rm erg~s^{-1}$ are employed in the calculation. See Section~\ref{sec:lep} for more discussion. }
\label{fig:case3}
\end{figure}
}

\section{Discussion and Summary}
To summarize, we calculated the diffuse gamma-ray intensity and neutrino intensity of the Galactic plane, which arise from hadronic interactions between CRs and the interstellar medium. We found that if we require the model neutrino intensity to be consistent with the 90\% C.L. upper limit of Galaxy, the model gamma-ray intensity is lower than the measured flux by the Tibet AS+MD array especially at the highest-energy bin in $398-1000\,$TeV. We speculated that an additional contribution by CRs with a relatively hard spectrum from the Cygnus region and/or some other extended sources of CRs beyond PeV is needed. {Provided that these sources can accelerate electrons to energy beyond 1\,PeV, the sub-PeV diffuse gamma-ray emissions may also contributed by the inverse Compton scatterings of PeV electrons off CMB, as long as the injected electron spectra are harder than $E^{-1.7}$ and the diffusion coefficient in the ambient source region is much larger than $0.015(B/5\mu\rm G)^2$ times the average diffusion coefficient of the interstellar medium. In such a scenario, Cygnus Cocoon would harbor one (or more) extreme particle accelerator of either protons or electrons} beyond PeV, and its gamma-ray spectrum might show a hardening above $30-100$\,TeV.
{If the sub-PeV gamma-ray flux from the Cygnus region turns out be low, other extreme particle accelerators in our Galaxy would make a more important contribution to the diffuse sub-TeV gamma-rays measured by the Tibet AS+MD array.}
The up-coming TeV -- PeV gamma-ray instrument LHAASO \citep{LHAASO19} is promising to test our speculation.

Lastly, we caveat that the upper limits of the neutrino flux employed in this work are only in 90\% C.L. Therefore, the constraint from the neutrino flux is not so strict and hence the contribution needed to be attributed to extended sources may be lower. {\citet{IC17_GP} showed a 2D likelihood scan of the normalization and spectral slope of Galactic neutrino flux. For the benchmark spectral index, i.e., 2.5, the 99\% C.L. flux upper limit is about 1.5 times the 90\% C.L. upper limit as employed in this study. Adopting the 99\% C.L. upper limit would relax the tension between the diffuse neutrinos and diffuse gamma rays.}
Also, the analysis for the neutrino upper limits of the  2HWC~J2031+415 and Gamma Cygni, the two sources related to the Cygnus region, assumes point-like source. A dedicated analysis optimized for the extended source could result in a different upper limit. Therefore, we encourage a detailed analysis of the neutrino emission on the Cygnus region with the spatial template of gamma-ray emission above 30\,TeV.

\section*{Acknowledgements}
We thank Christian Haack for helpful discussion on the IceCube data and the anonymous referee for the constructive comments. This work is supported by NSFC grants U2031105, 11625312 and 11851304, and the National Key R \& D program of China under the grant 2018YFA0404203.

\appendix
\section*{Diffusion and IC radiation of High-energy Electrons}
\begin{figure}
\centering
\includegraphics[width=0.8\textwidth]{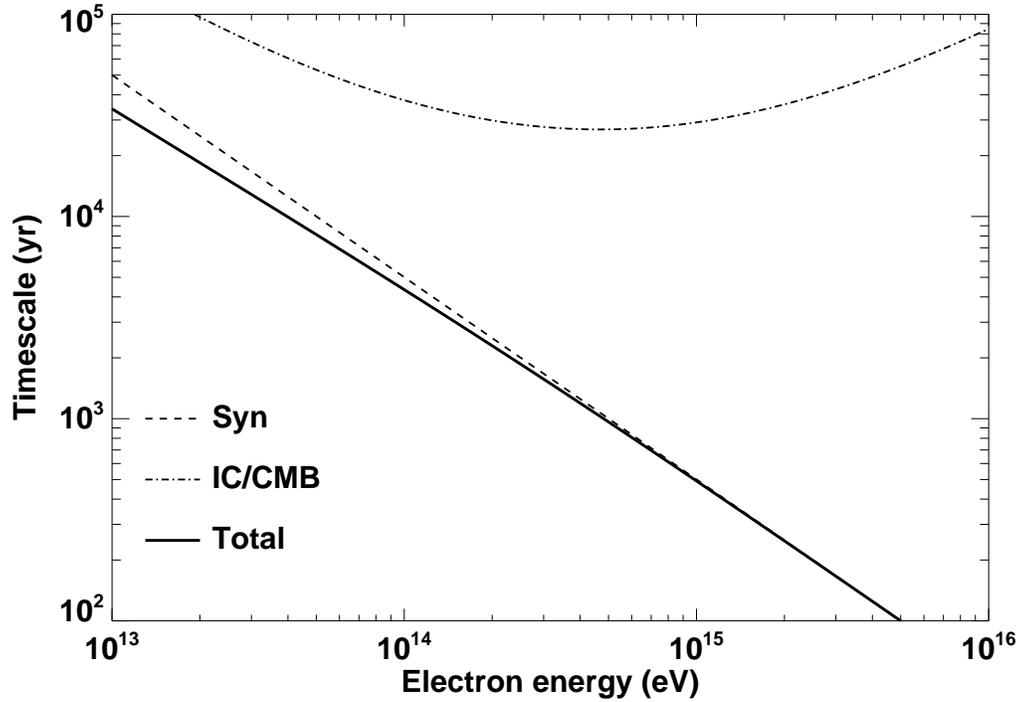}
\caption{Cooling timescale of electrons as a function of energy. The dashed line shows the synchrotron cooling timescale with $B=5\,\mu$G, the dash-dotted line shows the IC cooling timescale due to CMB radiation, and the solid line shows the cooling timescale for both two processes.}
\label{fig:tcool}
\end{figure}

We consider a simplified case in which high-energy electrons are injected at a constant rate at $r=0$ and diffuse isotropically to larger radius. The injection spectrum is given by $Q_{e,\rm inj}=Q_0E_e^{-p}$ for $E_e \leq 2\,$PeV. The prefactor $Q_0$ can be found by $\int E_eQ_{\rm inj}(E_e)dE_e=L_{e,\rm inj}$ with $L_{e,\rm inj}$ being the total electron luminosity, which will be treated as an input parameter. Following \citet{Syrovatsky59}, the electron distribution can be given by
 \begin{equation}\label{eq:diff_sol}
N(E_e,r)=\int_0^{t_{\rm age}} \frac{Q_{e,\rm inj}(E_g)dt}{(4\pi \lambda(E_e,t))^{3/2}}\exp\left[-\frac{r^2}{4\lambda(E_e,t)}\right]\frac{dE_g}{dE_e}
\end{equation}
where $t_{\rm age}\,$ is taken to be 1\,Myr and $\lambda(E_e,t)=\int_t^{t_{\rm age}}D(E_e'(t'))dt'$. $E_e'(t')$ represents the trajectory in the energy space of an electron the energy of which is $E_e$ at present, and $E_g$ is the initial energy of the electron at the generation (injection) time $t$. The relation between $E_e$ and $E_g$ as well as $dE_g/dE_e$ can be obtained by the energy evolution of the electron, i.e.,
\begin{equation}\label{eq:cooling}
\frac{dE_e}{dt}=-\frac{4}{3}\sigma_Tc\left(\frac{E_e}{m_ec^2}\right)^2\left[U_B+\Sigma_i U_{i,\rm ph}/\left(1+4\frac{E_e\epsilon_i}{m_e^2c^4}\right)^{3/2}\right]
\end{equation}
where $\sigma_T$ is the Thomson cross section, $m_e$ is the electron mass and $c$ is the speed of light. $U_B=B^2/8\pi$ is the magnetic field energy density. $U_{i,\rm ph}$ is the energy density of the $i$th radiation field. $\epsilon_i=2.82kT_i$ is the typical photon energy of the $i$th radiation field of a black body or a grey body radiation field with a temperature $T_i$ and $k$ is the Boltzmann constant \citep{Moderski05}. $U_{i,\rm ph}$ here includes the CMB radiation and the mean interstellar radiation field of the Galaxy which is composed of an infrared radiation field with $T=30\,$K and $U=0.3\,\rm eV~cm^{-3}$ and an optical radiation field with $T=5000\,$K and $U=0.3\,\rm eV~cm^{-3}$, although the interstellar radiation field has a negligible influence on PeV electrons due to the KN effect. The electron cooling timescale as a function of energy is shown in Fig.~\ref{fig:tcool}.

With the electron's spatial distribution, we can calculate the gamma-ray surface brightness profile $I_{\rm cyg}(E_\gamma,\theta)$ by integrating the IC radiation over the line of sight towards an arbitrary angle $\theta$ from the center of Cygnus Cocoon, following the method detailed in \citet{Liu19}. The total unmasked gamma-ray flux related to Cygnus Cocoon then reads 
\begin{equation}
F_{\rm cyg, \gamma}(E_\gamma)=2\pi\int_{0.5^\circ}^{\infty} I_{\rm cyg}(E_\gamma,\theta)\sin\theta d\theta.
\end{equation}
The contribution of extended sources in the leptonic case to the DGE can be obtained by Eq.~(\ref{eq:dge_source}).

\bibliography{ms}

\end{document}